\documentclass[%
 reprint,
 amsmath,amssymb,
 aps,prc,
]{revtex4-2}

\usepackage{graphicx}
\usepackage{dcolumn}
\usepackage{bm}
\usepackage{hyperref}
\usepackage{natbib}

\begin{document}


\title{Impact of $^7$Be breakup on $^7$Li(p,n) Neutron Spectrum}

\author{Midhun C.V$^1$}
\author{M.M Musthafa$^1$}\thanks{mmm@uoc.ac.in}%
\author{S.V Suryanarayana$^2$}
\author{T. Santhosh$^2$}
\author{A. Baishya$^2$}
\author{P. Patil$^3$}
\author{A Pal$^2$}
\author{P.C Rout$^2$}
\author{S Santra$^2$}
\author{R. Kujur$^2$}
\author{Antony Joseph$^1$}
\author{Shaima A$^1$}
\author{Hajara. K$^1$}
\author{P.T.M Shan$^1$}
\author{Satheesh B$^4$}
\author{Y. Sawant$^2$}

\author{B. V. John$^2$}
\author{E.T Mirgule$^2$}
\author{K.C Jagadeesan$^5$} 
\author{S. Ganesan$^6$} 

\affiliation{$^1$ Dept. of Physics,University of Calicut, Calicut University P.O Kerala, 673635 India}
\affiliation{$^2$ Nuclear Physics Division, Bhabha Atomic Research Centre, Mumbai 400085, India}
\affiliation{$^3$ Dept. of Physics, Karnatak University, Dharwad, Karnataka 580003, India}
\affiliation{$^4$ Dept. of Physics, Mahatma Gandhi Government Arts College, Mahe, 673311, India}
\affiliation{$^5$ Radiopharmaceuticals Division, Bhabha Atomic Research Centre, Mumbai 400085, India}
\affiliation{$^6$ Formarly Raja Ramanna Fellow, Bhabha Atomic Research Centre, Mumbai 400085, India}

\date{\today}

\begin{abstract}
The formation of continuum neutron distribution in $^7$Li(p,n) has been identified as due to the coupling of the $^7$Be breakup levels to the final state of the reaction. The continuum neutron spectra produced by $^7$Li(p,n) reaction has been estimated by measuring the double differential cross sections for continuum and resonant breakup of $^7$Be, through $^7$Li(p,n)$^7$Be$^*$ reaction at 21 MeV of proton energy. The breakup contributions from continuum and $5/2^-$, $7/2^-$ states of $^7$Be have been identified. The measured double differential cross sections have been reproduced through CDCC-CRC calculations. The cross sections were projected to neutron spectrum using Monte-Carlo approach and validated using experimentally measured $^3$He gated neutron spectra. $^7$Li(p,n) neutron spectrum at 20 MeV incident proton energy measured by McNaughton. et al. has been reproduced by adapting estimated model parameters for the reaction.


\end{abstract}

\maketitle


\section{Introduction}
Neutron induced reactions above 10 MeV are having a renewed interest as in the current scenario of fusion reactors and accelerator driven systems(ADS) \citep{ADSGanesan}. There, activation analysis is one of the promising method utilized to measure neutron induced cross sections, using quasi-monoenergetic neutron sources. Due to the energy tunability, and higher yield, the $^7$Li(p,n) channel is the most used accelerator based neutron source. Further, the $^7$Li(p,n) channel keeps the quasi-monoenergetic behaviour for a larger energy domain, upto 5 MeV, due to the excitation levels of $^7$Be being far enough (7/2$^{-}$ at 4.57 MeV) compared to the neutron energies, except the 1/2$^-$ level at 429keV\citep{EPEN}. 
\paragraph*{}As the proton energy increases further above 3.22 MeV, (1.64 MeV Threshold for $^7$Li(p,n) and 1.58 MeV breakup threshold for $^7$Be to $^3$He and $\alpha$) the neutron distribution behaves as a continuum from 0 to E$_P-$3.22 MeV, including the monoenergetic peaks corresponding to the population of ground and 1/2$^-$ states. However, there are several measurements with $^7$Li(p,n) neutrons, at proton energies above 15 MeV, for exploring the reactions having threshold greater than E$_p$ - 5MeV. This practice has been continued with the assumption that the neutrons contained in E$_p$ - 5MeV to E$_p$ - 3.2 MeV range is lesser than 1\% of neutrons enclosed by n$_0$ and n$_1$ (corresponding to ground and 1/2$^-$ states of $^7$Be) neutron colonies.  In some recent works, the extra contribution from the neutron continuum is being accounted  by a method of the tailing correction methods\citep{Siddharth,DLSmith}. 
\paragraph*{}So far  there exists only a single measurement of $^7$Li(p,n) neutron spectrum above 5 MeV, the domain which the neutron continuum distribution is being prominently contributing to the neutron spectrum. This measurement was performed by McNaughton et. al.\citep{MCNAUGHTON1975}, with a thick $^{nat}$Li target in the neutron time of flight mode, using the pulsed protons from cyclotron. There are some important works by S. G. Mashnik et. al. \citep{LAUR}, Meadows and Smith \citep{ANL} and Drosg et. al.\citep{DROSG2000}, to theoretically model the neutron spectrum from 7 Li(p,n) reaction,  by taking neutron spectrum measured by McNaughton et al. as a reference. However, due to the lack of enough experimental data on $^7\mathrm{Li}(p,n)^7\mathrm{Be^*}\rightarrow \mathrm{^3He}+\alpha$, the neutron spectrum is not well reproduced by these attempts.

\paragraph*{}In the present study, the neutron continuum distribution in the $^{7} \mathrm{Li}$(p,n) neutron spectrum, above three body breakup threshold, is considered to have emerged because of the coupling of continuum levels of $^7$Be to the outgoing neutron wave function. These continuum levels are being considered as originated by the relative motion of $\alpha-$ $^3$He internal structures of $^7$Be above the breakup threshold of 1.58 MeV. Moreover, other than continuum states, the 5/2$^-$, 7/2$^-$ resonant states of $^7$Be also contribute into the breakup. This makes the additional peak structures to the end region of the continuum neutron spectrum. This implies that the shape of continuum neutron spectrum has to be determined from the breakup cross section of $^7$Be at the continuum energies above breakup threshold. The continuum and resonant breakups of $^{7}\mathrm{Be}$ have been explored by D Chattopatyay et. al\citep{dipayan}, Summers and Nunes \citep{Summers}. Similarly, there important measurements from n\_TOF collaboration on $^7\mathrm{Be}(n,p)$ and $^7\mathrm{Be}(n,\alpha)$ channels, which show a similar coupling of $^3\mathrm{He} - \alpha$ breakup in the entrance channel \citep{7benp, 7bena}. However, in these works mentioned above, the entrance or exit conditions are  different from the requirements of the present work and therefore, these measured cross sections cannot be directly adapted for explaining three body continuum neutron spectrum from $^7$Li(p,n).
 
\paragraph*{}The measurement of $\alpha - ^3$He breakup cross sections from the continuum (resonant and non-resonant) states of $^7$Be requires the coincidence detection of $^3$He and $\alpha$, where the sum of break up threshold and relative energy between $^3He$ and $\alpha$ defines the excitation energy of $^7$Be. However due to the energy loss and energy and angular straggling of $^3$He and $\alpha$ in the target, make the reconstruction of excitation energies may sometimes become difficult. In a recent experiment using 1.2 mg/cm$^2$ Li target the same problem was observed due to the issues of straggling and energy loss in the target as well as thicker $\Delta$E detectors.\citep{faildkmat}.

\paragraph*{} 
Based on the analysis of data from previous attempt, the present experiment has been planned carefully and performed with a specially prepared thin $^{nat}$Li target, of 20$\mu$g/cm$^2$, sandwiched in between 5$\mu$g/cm$^2$ Al backing and 5$\mu$g/cm$^2$ carbon capping. It made the coincidence detection of $^3$He and $\alpha$ and reconstruction of $^7$Be excitation energy achievable due to a minimal energy and angular straggling. The double differential cross sections for $^7$Li(p,n)$^7$Be$^* \rightarrow$ $^3$He + $\alpha$ are measured and extended to the determination of neutron spectrum. The determined cross section has been verified with $^3$He gated neutron spectrum measured through time of flight technique. The details are presented in the following sections. 


\section{Experimental Setup} 
The experiment was performed at BARC-TIFR Pelletron-Linac facility, Mumbai, India, using  proton beam of 21 MeV. A specially fabricated 20$\mu$g/cm$^2$ $^{nat}$Li target with 5$\mu$g/cm$^2$ Al backing and 5$\mu$g/cm$^2$ carbon capping has been used for the experiment. The schematic of the experiment is illustrated in Fig. 1. Two silicon detector telescopes having 25$\mu$m and 1500$\mu$m $\Delta$E-E pairs, were mounted at $+$55$^{\circ}$ and $-$55$^{\circ}$ with a distance of 7.5 cm from the target, to record the ejectile particles.. Three EJ301 liquid organic scintillators having 12.7 cm diameter and 5 cm thick, are configured for measuring neutrons. Two EJ301 detectors are positioned at 30$^{\circ}$ and 60$^{\circ}$ at a distance of 1 meter from the target. The third detector was positioned at 30$^{\circ}$, 1.5 meter from the target. The n-$\gamma$ discrimination of the signal was performed using Mesytec MPD-4 module. Further the neutron detectors were configured in time of flight (TOF) with start from $\Delta$E detector signals. The telescopes were calibrated using $^{229}$Th $\alpha$ sources. Neutron detector thresholds were set using $^{137}$Cs sources and TDCs were calibrated using time calibrator module. 
\begin{figure}
\includegraphics[width=\columnwidth]{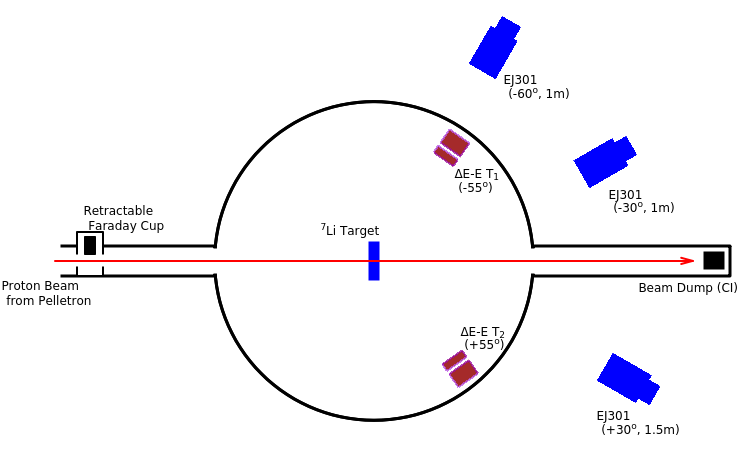}
\caption{Experimental Setup}
\end{figure}
\begin{figure}
\includegraphics[width=\columnwidth]{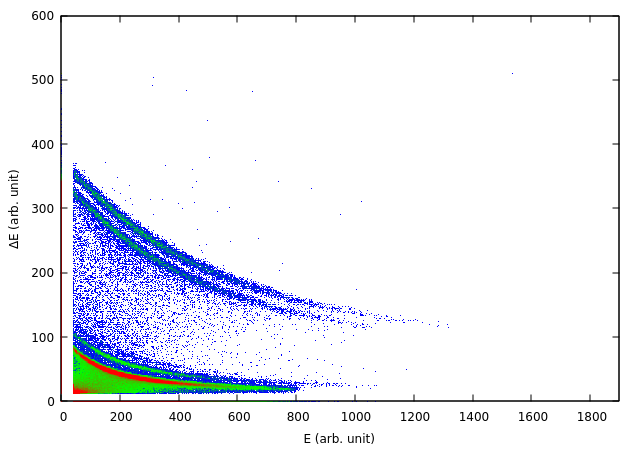}
\caption{$\Delta E - E$ correlation spectra measured by -55$^{\circ}$ telescope}
\end{figure}

\begin{figure}
\includegraphics[width=\columnwidth]{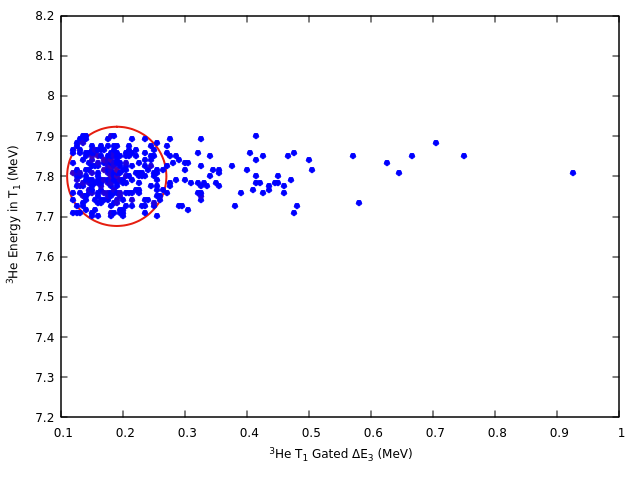}
\caption{Typical $\Delta E-$ $^3$He correlation with $\alpha$ fragment below particle discrimination threshold. The events enclosed in red circle represent the $\alpha$ events corresponding to identified $^3$He. Red circle is the boundary of the events around the kinematically defined query point.}
\end{figure}

\section{Data Analysis}

\subsection{Analysis of $^3$He$-\alpha$ coincidences }
The recorded events with E-$\Delta$E correlation is illustrated in Fig. 2. A banana gate on $^3$He band resolves the $\alpha$ particle events coincident with $^3$He. Some of the $\alpha$ particles produced from the reaction $^6$Li(p,$\alpha$)$^3$He. In addition meany of the random events are also possible. Hence the events corresponding to $^7$Li(p,n)$^7$Be$^* \rightarrow ^3$He+$\alpha$ reaction, which follow the thee body kinematics have been identified through fixed radius nearest neighbourhood approach. The kinematically allowed energies($E_{^3He}$ \& $E_{\alpha}$ ) corresponding to detector angles for the 21 MeV beam energy was taken as the quarry point. The constant radius is then considered as a window, and is taken as $\Delta = \sqrt{\sigma_{T1}^2+\sigma_{T2}^2}$, where $\sigma_{T1}$ and $\sigma_{T2}$ are the respective energy resolution given by the two telescopes. This accepts $^3$He$- \alpha$ coincidence events having both $E_{^3He}$ \& $E_{\alpha}$ that are inside the window. This makes the isolation of $^7$Li(p,n)$^7Be^* \rightarrow ^3$He+$\alpha$ events by removing $^6$Li(p,$\alpha$)$^3$He and other random coincidences. These type of events are addressed as 'true events'
\paragraph*{}There are events,  predicted by three body kinematics, having $E_{\alpha}$ below the particle discrimination threshold with the coincident $^3\mathrm{He}$ being above discrimination threshold, or vice versa. Such events were also reconstructed through the same technique. Thin gates were defined in the region of $^3$He/$^4$He band for which the kinematics predicts the possibility of counter particle below discrimination threshold. A correlation plot between energies in telescope 1 verses telescope 2, under this gate, has been generated. The quarry point was defined in the same approach adapted for 'true events'. Radius has been taken as $\Delta = \sqrt{{\sigma_{tel}}^2+ {\Delta_{gate}}^2 +{\sigma_{\Delta E}}^2 + E_{strag}}$, where ${\sigma_{tel}}$ is the energy resolution of the telescope which identified the particle. $\Delta_{gate}$ is the energy width of the gate applied on the particle band and ${\sigma_{\Delta E}}$ is the energy resolution of the $\Delta E$ detector which detects the counter particle. Further $E_{strag}$ is the energy straggling of the low energy events inside the target. $E_{strag}$ has been evaluated from SRIM code, for the particular target geometry, however found to be less compared to the $\Delta E$ resolution. The circle defined by the center as query point and radius $\Delta$ together forms a gate for counter-particle events and these events have been projected towards the $\Delta E$ energy. The centroid of the projected distribution is taken as the average energy of the counter particle. Such events will be addressed as reconstructed events. A typical $E_{^3\mathrm{He}}-\Delta$E correlation with defined circle for event selection is presented in Fig. 3.
\paragraph*{} The energy state of $^7$Be was reconstructed by adding $E_{^3He}$ and $E_{\alpha}$. The doubly differentiated cross sections for the laboratory folding angle between $^3$He $- \alpha$ = 110$^{\circ}$ and the neutron energy spectrum have been constructed through the energetics as,
\begin{equation}
E_n = E_{beam}-E_{\alpha}-E_{^3He}+E_{loss}+Q
\end{equation}
Where $E_n$ is the neutron energy, $E_{loss}$ is the energy loss of the particle inside the target and $Q$ is the Q-value for 3 body breakup (-3.23 MeV). 
\subsection{{\sc FRESCO} Calculations}
\paragraph*{} Continuum Discretized Coupled Channels (CDCC) calculations using {\sc fresco} \citep{Fresco} are performed for reproducing the measured double differential cross sections for $^7$Li(p,n)$^7$Be$^* \rightarrow ^3$He + $\alpha$. The energy distribution of emitted neutron is considered as arising due to the coupling of continuum states of $^7$Be to the outgoing neutron levels. The problem has been defined as p+$^7$Li (where, $^7$Li as $\alpha + t$ cluster ) for the entrance channel and $^7$Be+n (where, $^7$Be as $\alpha$ + $^3$He cluster) as exit channel mass partitions. The continuum states corresponding to the $^7$Be, above $\alpha-^3$He breakup threshold (1.59 MeV) have been generated and discretized with respect to the $\alpha-^3$He relative momentum. The relative momentum ($\hbar$k) upto 5 fm$^{-1}$ was considered for the analysis. This range was discretised into a number of bins in the interval of $\Delta$k=0.125fm$^{-1}$. The spin corresponding to each bin has been obtained as the vector sum of $\alpha - ^3$He relative angular momentum and spin of $^3$He. The optical potentials corresponding to p+$^7$Li have been taken from \citep{Pakou2,AP1,AP2,AP3,AP4} and the $\alpha$+n optical potentials have been obtained from \citep{SATCHLER}. The potentials corresponding to $^3$He+n have been obtained by fitting the angular distribution data reported by M. Drosg \citep{Drosg3He}. Differential cross section corresponding to each continuum bin has been generated through CDCC calculations. An alternative CDCC + CRC calculation was performed to account for the $^7$Be$^*$ $\rightarrow$ $^3$He+$\alpha$ angular distribution, including resonant breakup contribution from 7/2$^-$ and 5/2$^-$ levels. The overlaps from continuum states produced by $\alpha$+$^3$He and $^6$Li+p, above the breakup thresholds of 1.59 and 5.6 MeVs are considered for the estimation of breakup from the resonant levels. The spectroscopic amplitudes corresponding to continuum and resonant breakups have been adapted from the references, however the overlaps are different in those cases and the above, a fine optimization has been performed\citep{dipayan}. The calculated cross sections were cascaded to obtain the double differentiated cross sections corresponding to $^7$Li(p,n)$^7$Be$^* \rightarrow ^3$He + $\alpha$ reaction channel. The cascaded cross sections were projected for lab, corresponding to 110$^{\circ}$, via three body kinematics, to meet the experimental conditions. 

\subsection{Evaluation and Validation of Neutron Spectrum}
The full neutron spectrum has been generated through Monte-Carlo approach using the {\sc fresco} calculated cross sections, in  which the double differential part was matched with the experimental cross sections. This neutron spectrum has been validated through $^3$He gated neutron spectrum measured in the experiment. The $^3$He event gated neutron bands, identified from the TOF-PSD correlation plots corresponding to each neutron detectors were projected onto the TOF axis. The projected histogram has been converted to the energy spectrum by time calibration. Since the neutron spectrum is extending upto 16 MeV, the energy - efficiencies for neutron detectors were simulated using Geant4, upto 16 MeV, and compared with previously measured efficiencies \citep{Desai}. The energy-efficiency plot generated for EJ-301 neutron scintillators is presented in Fig. 4.

\paragraph*{}In order to  validate the theoretical estimates of neutron spectra based on the measured breakup cross sections, for 21MeV proton energy case, we proceeded as follows. The full neutron spectrum has been generated through  Monte-Carlo approach using a two step the procedure. \textbf{(a)} As described in section B, the CDCC+CRC calculations using FRESCO were carried out to obtain theoretical estimates of breakup cross sections and the double differential  cross sections which are  matched with the experimentally measured cross sections by the events in  telescopes. \textbf{(b)} Using the FRESCO results  of cross sections,  the neutron spectra corresponding to 30$^{\circ}$ and 60$^{\circ}$ neutron events which are in coincidence with $^3$He at 55$^{\circ}$, have been generated through Monte-Carlo approach. \textbf{(c)} From experimentally recorded  events in the EJ301 detectors,  we obtained $^3$He events at 55$^{\circ}$ gated neutron spectra. The validation of theoretical estimates by comparing \textbf{(b)} and \textbf{(c)} i.e., the Monte-Carlo generated with the experimentally measured neutron spectra. This procedure of validation has been carried out also for 20MeV proton energy. Since we did not measure for 20MeV proton energy, we proceed to use the experimental data reported  by Mc-Naughton et al using thicker Li target.  We performed a  set of FRESCO followed by Monte-Carlo calculations for 20 MeV  along with SRIM calculations to account the thick target effect  and these calculations reproduced  the spectrum reported by Mc-Naughton et al.,

\paragraph*{} Neutron spectra corresponding to 30$^{\circ}$ and 60$^{\circ}$ neutron events, in coincidence with 55$^{\circ}$ $^3$He, have been generated through Monte-Carlo approach based on {\sc fresco} calculated cross sections.  This spectra have been compared with the experimentally measured neutron spectra, for the validation of the theoretical estimates based on the measured breakup cross sections. Similarly, another set of {\sc fresco} and Monte-Carlo calculation has been performed for 20 MeV proton energy to reproduce the spectrum reported by McNaughton et al., along with SRIM \citep{SRIM} calculations to account the thick target effect. The n$_0$ and n$_1$ component to the neutron spectra have been calculated using {\sc fresco} including the ground (3/2$^-$)and 429 keV (1/2$^-$) states. {\sc fresco} calculated cross sections for  ground (3/2$^-$)and 429 keV (1/2$^-$) states were validated with ENDL-2019 evaluated cross sections for $^7$Li(p,n)$^7$Be\citep{tendl2019}. 

\begin{figure}
\includegraphics[width=\columnwidth]{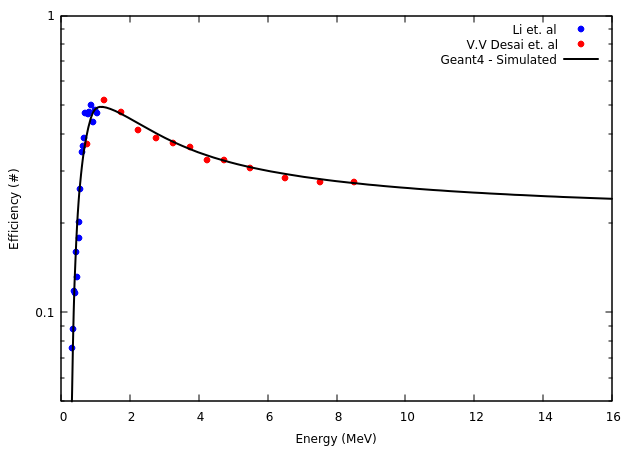}
\caption{Neutron Energy - Efficiency plot for EJ301 Liquid Scintillator}
\end{figure}

\section{Results and Discussion}

\begin{figure}
\includegraphics[width=\columnwidth]{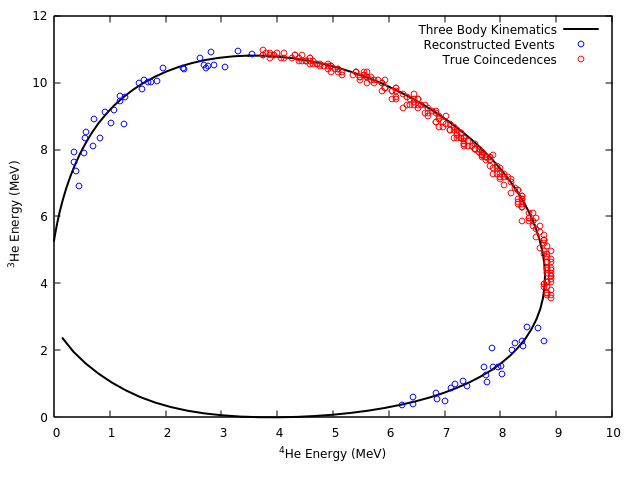}
\caption{The true coincidence and reconstructed events comparison with three body kinematics, projected to the fragments at $+55^{\circ}$ and $-55^{\circ}$}
\end{figure}

\begin{figure}
\includegraphics[width=\columnwidth]{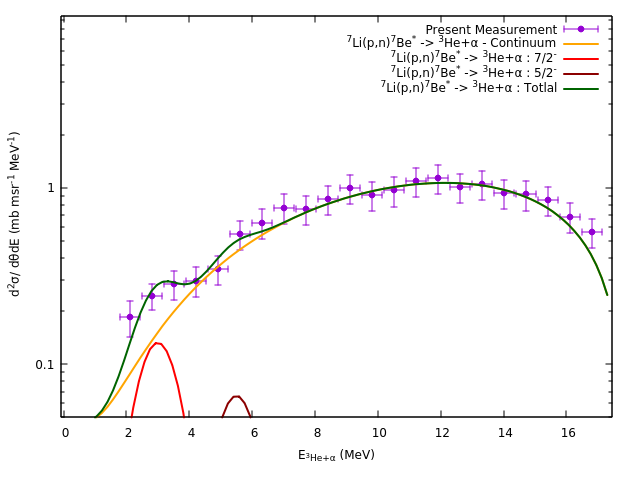}
\caption{The measured double differential cross sections for $^7$Li(p,n)$^7$Be$^*\rightarrow ^3$He+$\alpha$, and CDCC-CRC calculated continuum and resonant stare cross sections(for the laboratory folding angle of 110$^{\circ}$ )}
\end{figure}

Measured energy correlation for $^3\mathrm{He}-\alpha$ agrees well with the three body kinematics predicted for $^7$Li(p,n)$^7$Be$^*\rightarrow ^3$He+$\alpha$ reaction. Measurements are limited to $^3$He$-\alpha$ folding angle of 110$^{\circ}$ in laboratory frame, however, events got registered to a reasonable statistics. The $^3$He$-\alpha$ energy correlation, along with three-body kinematics is presented in Fig. 5. Measured double differential cross sections for the breakup corresponding to continuum states of $^7$Be are illustrated as locally averaged effective histogram values in Fig. 6, along with {\sc fresco}-CDCC calculations for $^7\mathrm{Li}(p,n)^7\mathrm{Be}^*\rightarrow^3$He+$\alpha$. The comparison of measured double differential cross section with CDCC-CRC results shows a good agreement and illustrates the physics of neutron continuum formation in this typical reaction channel. Based on CDCC-CRC analysis, it is well understood that the continuum neutron distribution is formed due to the coupling of breakup levels to the $^7$Be + n state. Other than the continuum, there exists 7/2$^-$ and 5/2$^-$ levels above breakup threshold, which contribute as the resonant breakups. In the present study, breakup contributions from these resonant levels are observed with a lesser statistics. However, these peaks are being reproduced with CDCC-CRC calculations. The spectroscopic amplitudes (SA) of 0.26, 1 and 0.57 have been used in reproducing continuum, $7/2^-$ and $5/2^-$ contributions respectively.
 \paragraph*{}The Monte-Carlo based calculation for 0$^{\circ}$ neutrons has been performed using CDCC-CRC calculated cross section, with parameters obtained through matching the experimentally measured double differential cross sections. The CRC were considered for accounting the resonant breakup contribution. The estimated $0^{\circ}$ neutron spectrum, corresponding to the breakup is presented in Fig. 7. This shows the contribution from 7/2$^-$(4.57 MeV) and 5/2$^-$ (6.73 and 7.21 MeV) levels as  Gaussian peaks. Due to the overlaps of the continuum levels to the resonant states, the width of the resonant breakup neutron colonies is large.

\begin{figure}
\includegraphics[width=\columnwidth]{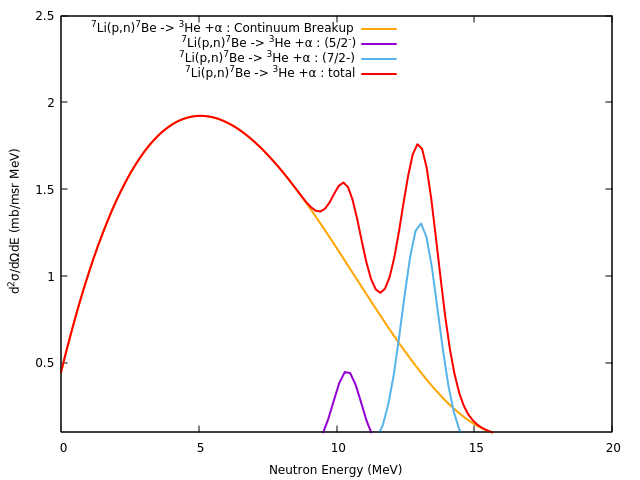}
\caption{The breakup neutron spectrum, having continuum and resonance contributions}
\end{figure}

\begin{figure}
\includegraphics[width=\columnwidth]{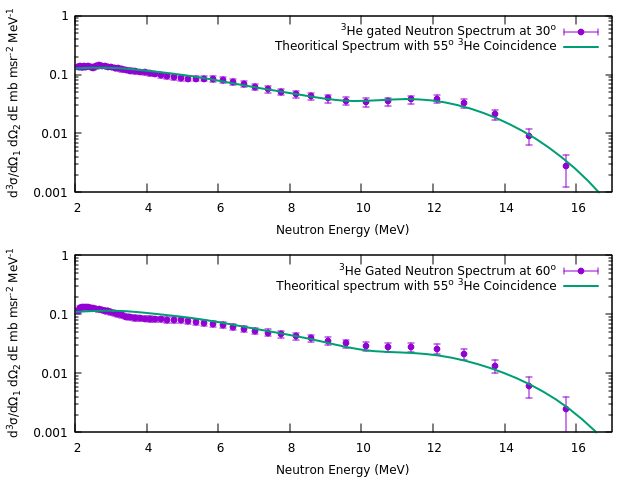}
\caption{The comparison of $^3$He gated neutron spectra with theoretical spectra}
\end{figure}
\begin{figure} 
\includegraphics[width=\columnwidth]{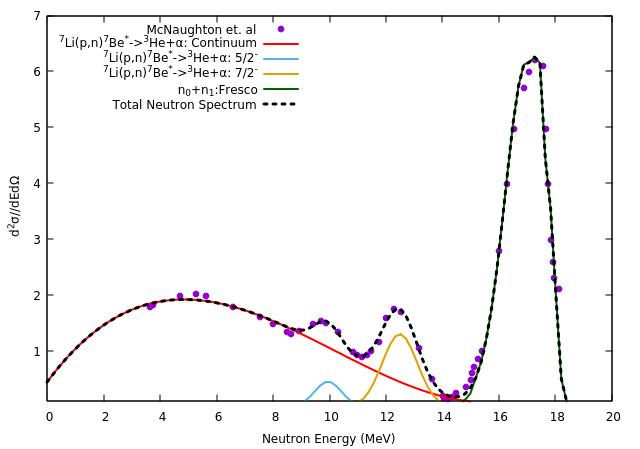}
\caption{Comparison of evaluated neutron spectrum with the spectrum reported by McNaughton et. al.}
\end{figure}
\paragraph*{}The validation of theoretically calculated neutron spectra has been performed with experimentally measured 55$^{\circ}$ $-$ $^3$He gated neutron spectrum at 30$^{\circ}$ and 60$^{\circ}$. The theoretical spectra for these angles are generated through Monte-Carlo approach over the theoretical cross sections, as described above. The comparison of theoretically evaluated neutron spectrum with 55$^{\circ}$ $-$ $^3$He gated neutron spectrum at 30$^{\circ}$ and 60$^{\circ}$ is presented in Fig. 8.  The comparison shows that the theoretically generated neutron spectra are well reproducing the experimental neutron spectrum within the counting statistics of the neutrons per energy bin.

\paragraph*{}The comparison of theoretical neutron spectra for 20 MeV of proton energy and the neutron spectra reported by McNaughton et al. is illustrated in Fig. 9. This shows the theoretical calculations, with parameters used for reproducing the double differential cross section of $^7$Li(p,n)$^7$Be$^*\rightarrow ^3$He+$\alpha$ are quite apt for generating the breakup neutron spectrum from p + $^7$Li system. The $n_0$ and $n_1$ neutron colonies are well reproduced by the {\sc fresco} calculations, optimized and validated with respect to TENDL-2019 evaluations.

\paragraph*{}The continuum neutron colony, emerges due to the coupling of continuum levels generated by the $\alpha + ^3$He cluster structure to the $^7$Be+n final state, has been qualitatively measured by the present study. Effect of coupling of $\alpha + ^3$He continuum levels makes a neutron distribution from 0 MeV to E$_{beam}$ - Q, where Q is the sum of Q-value for $^7$Li(p,n) and the breakup threshold of $^7$Be to $\alpha$ and $^3$He. Further, at the high energy tail of the breakup continuum, there form two additional peaks, corresponding to the 7/2$^-$ and 5/2$^-$ resonance levels of $^7$Be, above the breakup threshold. The overlaps of breakup levels of $^7$Be produced by $^3$He +$\alpha$ and $^6$Li + p structures, above 1.59 and 5.60 MeVs, produce broadening of the resonance neuron colonies. However for the practical use, the neutron distribution at the lower energy will be different, due to the presence of satellite neutrons as recommended by Drosg evaluations\citep{DROSG2000}. The satellite neutron distributions are formed through multiple scattering of neutrons, which are highly dependent on the experimental setup. However, the present study is concluded without accounting the satellite neutrons, formed at lower energies. 
\section{Summary}
The formation of the continuous neutron distribution for $^7$Li(p,n) reaction at higher proton energies has been identified as the overlap of continuum levels generated by $\alpha +^3$He clusters of $^7$Be above breakup threshold. This has been verified by measuring the double differential cross sections for three body breakup at 21 MeV of proton energy. The measured double differential cross sections have been reproduced through {\sc fresco} calculations by adding the coupling of continuum generated by $\alpha +^3$He to the $^7$Be+n final state. The neutron spectrum has been simulated through Monte-Carlo with using {\sc fresco} calculated cross sections. These neutron spectra have been validated using experimentally measured $^3$He gated spectrum and McNaughton's spectrum reported for 20 MeV.
\section{Acknowledgements}
The support of BARC-TIFR Pelletron-LINAC group during the experiment is acknowledged.The authors acknowledge the staff in TIFR Target Laboratory in the development of sandwiched Li target for the experiment. The authors acknowledge, A.K Bhakshi and Rupali Pal, RP$\&$AD BARC, Mumbai and A. Shanbhag, Health Physics Division, BARC for their kind support during the experiment. This study is part of the project supported by DAE-BRNS, Sanction Order 36(6)/14/30/2017-BRNS/36204.

\bibliographystyle{unsrt}
\bibliography{Refer}

\begin{thebibliography}{10}

\bibitem{ADSGanesan}
S.~Ganesan.
\newblock Nuclear data requirements for accelerator driven sub-critical systems
  – a roadmap in the indian context.
\newblock {\em Pramana - J Phys}, 68:257--268, 2007.

\bibitem{EPEN}
Rebecca Pachuau, B.~Lalremruata, N.~Otuka, L.~R. Hlondo, L.~R.~M. Punte, and
  H.~H. Thanga.
\newblock Thick and thin target $^7$Li(p,n)$^7$Be neutron spectra below the
  three-body breakup reaction threshold.
\newblock {\em Nuclear Science and Engineering}, 187(1):70--80, 2017.

\bibitem{Siddharth}
{Parashari, Siddharth}, {Mukherjee, S.}, {Naik, H.}, {Suryanarayana, S. V.},
  {Makwana, Rajnikant}, {Nayak, B. K.}, and {Singh, N. L.}
\newblock Measurement of the $^{58}\mathrm{Ni}$(n, p)$^{58}\mathrm{Co}$ and $^{58}\mathrm{Ni}$(n,2n)$^{57}\mathrm{Ni}$ reaction
  cross-sections for fast neutron energies up to 18 MeV.
\newblock {\em Eur. Phys. J. A}, 55(4):51, 2019.

\bibitem{DLSmith}
D.L. Smith, Plompen, and A.J.M. Semkova.
\newblock Corrections for low energy neutrons by spectral indexing.
\newblock {\em OECD NEA/WPEC-19}, 19, 2005.

\bibitem{MCNAUGHTON1975}
M.W. McNaughton, N.S.P. King, F.P. Brady, J.L. Romero, and T.S. Subramanian.
\newblock Measurements of $^7$Li(p,n) and $^9$Be(p,n) cross sections at 15, 20 and 30
  MeV.
\newblock {\em Nuclear Instruments and Methods}, 130(2):555 -- 557, 1975.

\bibitem{LAUR}
S.~G. Mashnik, M.~B. Chadwick, P.~G. Young, R.~E. MacFarlane, and L.~S. Waters.
\newblock $^{7}\mathrm{Li}$(p,n) nuclear data library for incident proton
  energies to 150 MeV.
\newblock Technical Report LA-UR-00-1067, Los Alamos National Laboratory, 2000.

\bibitem{ANL}
J.W Meadows and D.L Smith.
\newblock Neutrons from proton bombardment of natural lithium.
\newblock Technical Report ANL-7938, Argone National Laboratory, 1972.

\bibitem{DROSG2000}
M.~Drosg.
\newblock Sources of variable energy monoenergetic neutrons for fusion-related
  applications.
\newblock {\em Nuclear Science and Engineering}, 106(3):279--295, 1990.

\bibitem{dipayan}
D.~Chattopadhyay, S.~Santra, A.~Pal, A.~Kundu, K.~Ramachandran, R.~Tripathi,
  T.~N. Nag, and S.~Kailas.
\newblock Direct and resonant breakup of radioactive $^{7}\mathrm{Be}$ nuclei
  produced in the $^{112}\mathrm{Sn}(^{6}\mathrm{Li},^{7}\mathrm{Be})$
  reaction.
\newblock {\em Phys. Rev. C}, 102:021601, Aug 2020.

\bibitem{Summers}
N.~C. Summers and F.~M. Nunes.
\newblock $^{7}\mathrm{Be}$ breakup on heavy and light targets.
\newblock {\em Phys. Rev. C}, 70:011602, Jul 2004.

\bibitem{7benp}
L.~M. Damone et al.
\newblock $^{7}\mathrm{Be}(n,p)^{7}\mathrm{Li}$ reaction and the cosmological
  lithium problem: Measurement of the cross section in a wide energy range at
  n\_TOF at CERN.
\newblock {\em Phys. Rev. Lett.}, 121:042701, Jul 2018.

\bibitem{7bena}
M.~Barbagallo et al.
\newblock $^{7}\mathrm{Be}(n,\ensuremath{\alpha})^{4}\mathrm{He}$ reaction and
  the cosmological lithium problem: Measurement of the cross section in a wide
  energy range at n\_tof at cern.
\newblock {\em Phys. Rev. Lett.}, 117:152701, Oct 2016.

\bibitem{faildkmat}
\newblock Private communications, data received from S.V Suryanarayana, E.T. Mirgulae  et. al.,
\newblock This data has been reanalysed by our group recently.

\bibitem{Fresco}
Ian~J. Thompson.
\newblock Coupled reaction channels calculations in nuclear physics.
\newblock {\em Computer Physics Reports}, 7(4):167 -- 212, 1988.

\bibitem{Pakou2}
A.~Pakou et al.
\newblock Probing the cluster structure of $^{7}\mathrm{Li}$ via elastic
  scattering on protons and deuterons in inverse kinematics.
\newblock {\em Phys. Rev. C}, 94:014604, Jul 2016.

\bibitem{AP1}
George Freier, Eugene Lampi, W.~Sleator, and J.~H. Williams.
\newblock Angular distribution of 1- to 3.5-MeV protons scattered by
  ${^{4}\mathrm{He}}$.
\newblock {\em Phys. Rev.}, 75:1345--1347, May 1949.

\bibitem{AP2}
Philip~D. Miller and G.~C. Phillips.
\newblock Scattering of protons from helium and level parameters in
  ${^{5}\mathrm{Li}}$.
\newblock {\em Phys. Rev.}, 112:2043--2047, Dec 1958.

\bibitem{AP3}
J.~E. Brolley, T.~M. Putnam, L.~Rosen, and L.~Stewart.
\newblock Hydrogen-helium isotope elastic scattering processes at intermediate
  energies.
\newblock {\em Phys. Rev.}, 117:1307--1316, Mar 1960.

\bibitem{AP4}
R.~Kankowsky, J.C. Fritz, K.~Kilian, A.~Neufert, and D.~Fick.
\newblock Elastic scattering of polarized protons on tritons between 4 and 12
  MeV.
\newblock {\em Nuclear Physics A}, 263(1):29 -- 46, 1976.

\bibitem{SATCHLER}
G.R. Satchler, L.W. Owen, A.J. Elwyn, G.L. Morgan, and R.L. Walter.
\newblock An optical model for the scattering of nucleons from $^4$He at energies
  below 20 MeV.
\newblock {\em Nuclear Physics A}, 112(1):1 -- 31, 1968.

\bibitem{Drosg3He}
M.~Drosg, R.~Avalos Ortiz, and P.~W. Lisowski.
\newblock Neutron interactions with $^{3}\mathrm{He}$ revisited—i: Elastic scattering
  around and beyond 10 MeV.
\newblock {\em Nuclear Science and Engineering}, 172(1):87--101, 2012.

\bibitem{Desai}
V.~V. Desai, B.~K. Nayak, A.~Saxena, S.~V. Suryanarayana, and R.~Capote.
\newblock Prompt fission neutron spectra in fast-neutron-induced fission of
  $^{238}\mathrm{U}$.
\newblock {\em Phys. Rev. C}, 92:014609, Jul 2015.

\bibitem{SRIM}
James~F. Ziegler.
\newblock Srim-2003.
\newblock {\em Nuclear Instruments and Methods in Physics Research Section B:
  Beam Interactions with Materials and Atoms}, 219-220:1027 -- 1036, 2004.
\newblock Proceedings of the Sixteenth International Conference on Ion Beam
  Analysis.


\bibitem{tendl2019}
A.J. Koning, D.~Rochman, J.-Ch. Sublet, N.~Dzysiuk, M.~Fleming, and S.~{van der
  Marck}.
\newblock Tendl: Complete nuclear data library for innovative nuclear science
  and technology.
\newblock {\em Nuclear Data Sheets}, 155:1 -- 55, 2019.
\newblock Special Issue on Nuclear Reaction Data.

\end{thebibliography}
\end{document}